\documentclass{aa}  

\usepackage{graphicx}
\usepackage{orcidlink}
\usepackage{txfonts}
\usepackage{mathtools}
\begin{document}

   \title{Improving Monte Carlo radiative transfer simulations:\\ A shift of framework }
    \titlerunning{Shift of framework}

   \author{Anton~Krieger \orcidlink{0000-0002-3639-2435}
          \and
          Sebastian~Wolf \orcidlink{0000-0001-7841-3452}
          }

   \institute{Institut für Theoretische Physik und Astrophysik, Christian-Albrechts-Universität zu Kiel, Leibnizstra{\ss}e 15, 24118 Kiel, Germany.\\
             Corresponding \email{akrieger@astrophysik.uni-kiel.de}
             }

   \date{Received 9 June 2023 / Accepted 26 October 2023}

 
  \abstract
   {
    Monte Carlo radiative transfer (MCRT) simulations are a powerful tool for determining the appearance of astrophysical objects, analyzing the prevalent physical conditions within them, and inferring their properties on the basis of real observations. Consequently, a broad variety of codes has been implemented and optimized with the goal of solving this task efficiently. To that end, two distinct frameworks have emerged, namely, the extinction and the scattering framework, which form the basis of the path determination procedures of those codes. These procedures affect the step length of simulated photon packages and are used for determining flux estimates. Despite the fact that these simulations play an important role at present and thus require significant computational resources, little attention has been paid to the benefits and the drawbacks of both frameworks so far. In this study, we investigate their differences and assess their performance with regard to the quality of thereby obtained flux estimates, with a particular focus on the required computational demand. To that end, we use a testbed composed of an infinite plane-parallel slab, illuminated from one side, and we determine transmitted intensity using MCRT simulations for both frameworks. We find that there are vast differences between the frameworks with regard to their convergence speed. The scattering framework outperforms the extinction framework across all considered optical depths and albedos when solving this task, particularly in the regime of high optical depths. Its implementation can therefore greatly benefit all modern MCRT codes as it has the potential to significantly reduce required computation times. Thus, we highly recommend its consideration for various tasks that require MCRT simulations.
   }


   \maketitle
%

\section{Introduction}

    Radiative transfer (RT) simulations provide a crucial tool in astrophysics across various areas of application. These simulations allow us to predict and analyze the wavelength-dependent appearance of astrophysical objects, linking it to the underlying physical properties of these systems (e.g., protoplanetary disks, exoplanet atmospheres, filaments). To enable the simulation of complex environments, modern RT simulations often apply the Monte Carlo (MC) approach, which relies on the probabilistic simulation of so-called photon packages (PPs) and their randomly generated paths through a model space. Within the last two decades, various MCRT codes have been developed, including MC3D \citep{1999A&A...349..839W,Wolf_2003}, MCFOST \citep{2006A&A...459..797P,2022ascl.soft07023P}, MCMax \citep{2009A&A...497..155M}, RADMC-3D \citep{2012ascl.soft02015D}, Mol3D \citep{Ober_2015}, POLARIS \citep{2016A&A...593A..87R}, and SKIRT\,9 \citep{2020A&C....3100381C}. These codes have been optimized with regard to their computation time by using such methods as a locally divergence free continuous absorption \citep{1999A&A...344..282L}, immediate reemission scheme according to a temperature corrected emission spectrum \citep{2001ApJ...554..615B}, partial diffusion approximation \citep{2009A&A...497..155M}, modified random walk \citep{2009A&A...497..155M,2010A&A...520A..70R}, biasing techniques \citep{2016A&A...590A..55B}, precalculated sphere spectra \citep{2020A&A...635A.148K}, and an extended peel-off method \citep{2021A&A...645A.143K}. 
    
    Nonetheless, the analyses adopting these methods may remain limited due to a lack of required computational resources, in particular, the computation time. This is especially the case for the simulations of the RT in systems of high optical depth, for which we have to consider high numbers of simulated interactions a PP undergoes before leaving the model space and, consequently, the computation time rises. For such systems, it has been found that an overly low number of simulated PPs or an insufficient number of simulated interactions may even lead to the so-called scattering order problem \citep{2021A&A...645A.143K},  resulting in severely underestimated flux values \citep[e.g.,][]{2016A&A...590A..55B,2017A&A...603A.114G,2018ApJ...861...80C,2020A&A...635A.148K,2021A&A...645A.143K}. We note that the underestimation is a consequence of the needed restriction of simulated PPs and a consequence of very high MC noise combined with non-Gaussian statistics. In that regard, the scattering order problem generally states that the calculation of a reliable flux estimate requires a proper representation of simulated scattering orders, with the latter being the number of interactions a PP has undergone prior to its detection. However, the corresponding scattering order distribution has been shown to widen and shift toward larger scattering orders as the transverse optical depth of a system increases, quickly leading to infeasible demands for MCRT simulations and computational limits that can be difficult to overcome. 

    Additionally, the problem of unreliable flux estimates has also been reported for systems with optical depths between 10 and 30 \citep{2017A&A...603A.114G}. As a result, it can even affect simulations of embedded radiation sources, such as young accreting planets \citep{2022A&A...662A..99K}. Moreover, three-dimensional (3D) simulations of fainter sources can be extremely challenging, if the number of simulated sources is high and the feasible number of simulated PPs per source is small and becomes a limiting factor. In particular, the simulation of flux and polarization maps of protoplanetary disks resulting from self-scattering (i.e., dust grains thermally emitting photons that scatter off other dust grains) is a computationally extremely demanding task. However, its effect can be crucial in terms of the inference of disk properties based on observations \citep[e.g.,][]{2020A&A...640A.122B,2021A&A...648A..87B}.

    Therefore, it is generally highly desirable to develop the tools and understanding that would allow us to expand the scope and complexity of simulated objects, while maintaining the expected reliability of the results. Motivated by this goal, we explore the difference of two fundamental MCRT frameworks. Even though both frameworks are generally well known, we find that the most popular MCRT codes currently stick to either one of them with no consideration of the other. We show that these frameworks, however, may differ significantly with regard to their required computation speed and quality of flux estimates. To that end, we use a setup composed of an infinite plane-parallel slab, illuminated from one side, as a testbed, and we compare the results of the MCRT simulations of the transmitted intensity performed in both frameworks. In Sect. \ref{sec:methods}, we briefly summarize the frameworks and introduce the setup. Subsequently, we perform MCRT simulations to compare the performance of the frameworks in Sect. \ref{sec:results}. Lastly, we discuss implications regarding the choice of framework depending on the albedo and optical depth of the system in Sect. \ref{sec:discussion_and_conclusion}, before presenting our final conclusions.

\section{Methods}
\label{sec:methods}
    In this section, we describe two frameworks, which we call the extinction framework (EF) and the scattering framework (SF), which are both widely used in modern MCRT codes when estimating flux values. The underlying procedure of both frameworks is based on the simulation of PPs on randomly generated paths, which follow certain probability density functions. On its path, a PP may interact multiple times with the medium before eventually leaving the model space. The interactions usually include absorption and scattering, both of which are a cause for extinction. As a result, the weight carried by a PP, which corresponds to an intensity or energy, may be reduced prior to its detection by an observer. To properly simulate this process, the optical properties of the interacting medium have to be taken into account. These include the cross-sections for scattering, $C_{\rm sca}$, absorption, $C_{\rm abs}$, and extinction, $C_{\rm ext} = C_{\rm sca} + C_{\rm abs}$, as well as the (single scattering) albedo, $A=C_{\rm sca}/C_{\rm ext}$. These quantities additionally determine the optical depth $\tau$ along a straight path of length $\Delta l$ through the medium with $\tau = C_{\rm int} \rho \Delta l $, where $C_{\rm int}$ is either of the aforementioned cross-sections and $\rho$ the (average) number density along the path. During a MCRT simulation, the path lengths between two consecutive events of interaction are determined on the basis of randomly chosen optical depths, which follow an exponential distribution. 

\subsection{MCRT frameworks}
\label{sec:frameworks}
    The considered frameworks differ with regard to the procedure of path determination. In the EF, the path length between two consecutive events of interaction is determined on the basis of extinction optical depths, $\tau_{\rm ext}$, and the weight of an interacting PP is reduced at each point of interaction by a factor that equals the albedo of the medium. At its core, this procedure first randomly selects a location for the subsequent interaction event and then selects its interaction type. For improved performance, the PP is split at the location of interaction into an absorbed and a scattered part, of which only the latter is traced during the flux determination, carrying a weight that is reduced by a weighting factor that corresponds to the albedo, $A$. In the SF, on the contrary, the process of absorption is assumed to occur continuously along the PP path, which can be interpreted as leaving out the probabilistic determination of the absorption process. In other words, this framework allows us to ``passify'' the absorption process. As a consequence, only the scattering process is actively simulated in a probabilistic manner, such that the scattering optical depth is used as a basis to determine the path length between two consecutive scattering events and the weight is reduced passively. We note that MCRT simulations, which contrary to this study trace the absorbed radiation energy, have been shown to significantly benefit from the usage of a continuous absorption procedure \citep{1999A&A...344..282L}. However, throughout this study, we only consider the effect of its choice on derived flux estimates. 

    In particular, the implementation of these frameworks encompasses three aspects. First, there is the determination of a pseudo random number $r\in \left[0,\,1\right)$ according to a uniform distribution. Second, there is the calculation of the (corresponding) optical depth, $\tau=-\ln{\left(1-r\right),}$ and, third, the framework-dependent determination of the thus resulting interaction-free length, $\Delta l$, and weighting factor, $w$. In the EF (SF), the latter are given by $\Delta l = \tau/\rho/C_{\rm ext}$ ($\Delta l = \tau/\rho/C_{\rm sca}$) and $w=A$ ($w=e^{-\Delta l \rho C_{\rm abs}}$). We note that these equations only apply if the PP experiences another interaction at its subsequent location of interaction. If, on the contrary, it leaves the interacting medium after traversing a distance of only $\Delta l'<\Delta l$, the weighting factor is not applied within the EF and a changed weighting factor of $w=e^{-\Delta l' \rho C_{\rm abs}}$ is instead used in the SF. This procedure ensures that the PP leaves the medium with a proper weight. 

    In the limit of infinite simulated PPs, both frameworks yield the same result in terms of estimated flux values. However, their differences may show in the form of:\ 1) different convergence speeds in terms of the number of required PPs or simulation time;  2) the computation time per PP; and 3) the level of variance of flux estimates (MC noise) as a function of the number of simulated PPs.

    We note, that (in principle) there are many more conceivable frameworks that could be used, however, the EF and SF are the most prominent. Moreover, these frameworks can be transformed into each other by introducing the correct constant stretching factor. As a result of this biasing procedure, the weighting factor would be affected, as the path determination and weighting scheme are interdependent. Eventually, whether the photons are removed along the photon path (SF) or only at the location of the scattering event (EF) depends solely on the chosen stretching factor. A brief derivation of the transformation that links both frameworks is presented in Sect. \ref{sec:appendix:EF2SF} in the appendix.

\subsection{Setup}
\label{sec:setup}
    To explore the differences of these frameworks, we adopted the setup of \citet{2018ApJ...861...80C}, which is composed of an infinite plane-parallel 3D homogeneous slab with a total transverse (extinction) optical depth of $\tau_{\rm max}$, which is embedded in a vacuum. This setup is often used as a testbed for MCRT simulations, in which the slab is illuminated by an isotropic (monochromatic) radiation source from one side and the intensity transmitted through the slab is measured, as a function of $\mu = \cos{\theta}$, by a simulated observer on the other side of the slab \citep[for details regarding the setup, see][]{2018ApJ...861...80C,2021A&A...645A.143K}. Here, the quantity $\theta$ describes the penetration angle of the transmitted PPs with $\theta = 0$ corresponding to the direction perpendicular to the surface of the slab. In total, we use $M_{\rm bin}=41$ detector bins, which are linearly sampled regarding the direction $\mu\in \left[0,1 \right]$. The transmitted intensity of the $j$-th detector bin is then given by:
    \begin{equation*}
        I_{\rm trans}(\mu_j) = \sum_{i=1}^{N} \frac{w_{i,\,j}}{N} \frac{M}{2\mu_j},
    \end{equation*}
    where $N$ is the number of simulated PPs and $w_{i,\,j}$ is the total weight that the $i$th PP contributes to the detector bin, corresponding to the direction $\mu_j$. Unless mentioned otherwise, we assume an albedo of $A=0.5$ and simulate scattering events using an isotropic phase function. For this setup, the result for the transmitted intensity can also be calculated on the basis of a non-probabilistic method \citep[see][]{2018ApJ...861...80C,2021A&A...645A.143K}, which we use to assess the quality of the obtained MCRT-based results.

\section{Results}
\label{sec:results}

   \begin{figure}
   \centering
   \includegraphics[width=\hsize]{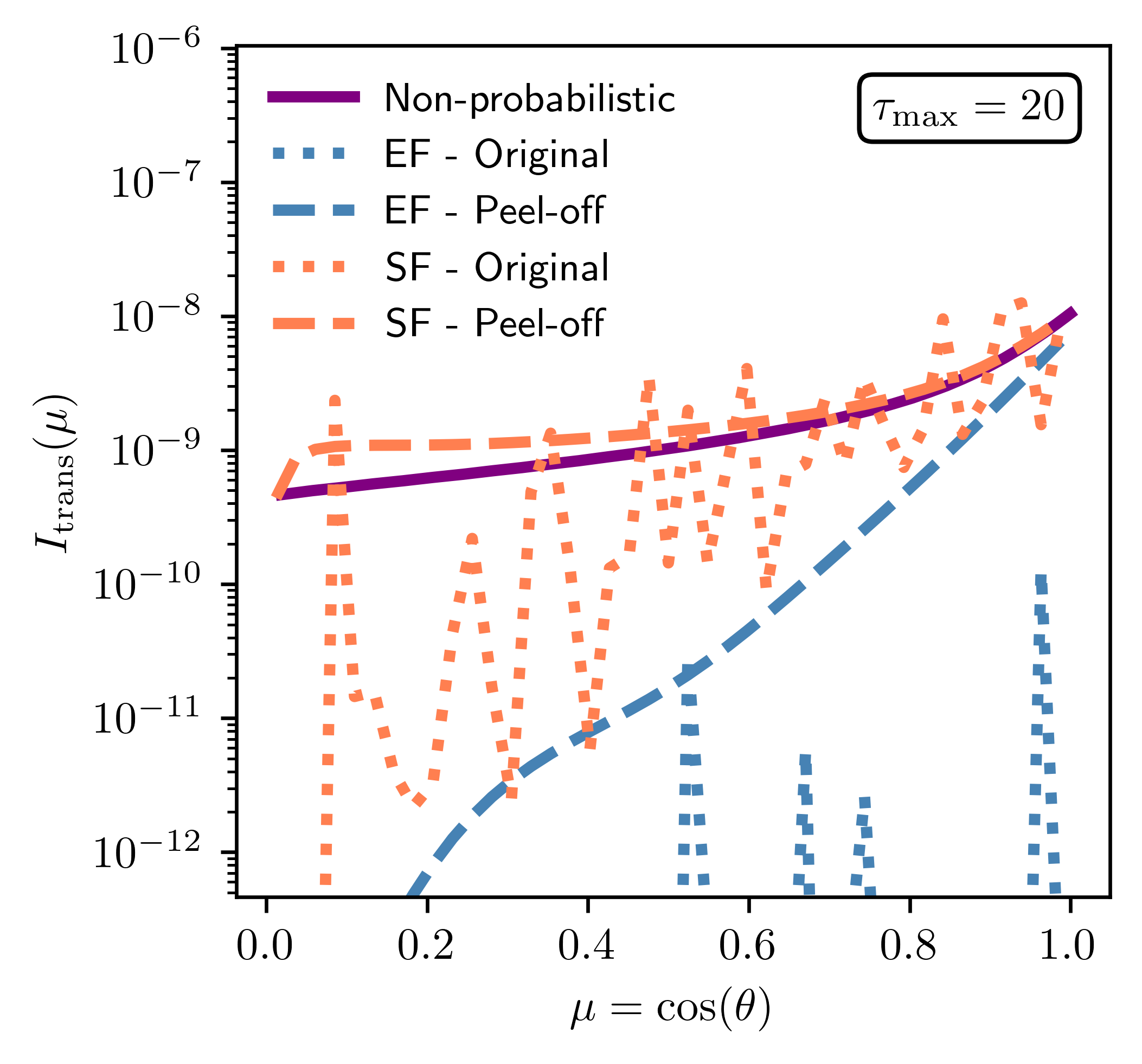}
      \caption{Transmission curves for $\tau_{\rm max}=20$ using $N_{\rm trans}=10^4$ transmitted PPs per MC-based simulation.}
         \label{fig:transmission_tau20}
   \end{figure}

    To compare the performance of the SF with that of the EF (see Sect. \ref{sec:frameworks}), we performed MCRT simulations to obtain the transmitted intensity for the slab setup (see Sect. \ref{sec:setup}). We assumed an (extinction) optical depth of $\tau_{\rm max}=20$ and ran each simulation until a total of $N_{\rm trans}=10^4$ PPs were successfully transmitted through the slab. Consequently, the required number of simulated PPs $N$ for both frameworks may differ. For each of these simulations, we additionally calculated the corresponding transmission curve that would emerge if the peel-off method \citep{1984ApJ...278..186Y} was applied, which is a commonly used MCRT method, that has shown to significantly boost the performance of MCRT simulations \citep[e.g.,][]{2021A&A...645A.143K}. The obtained transmission intensity is shown in Fig. \ref{fig:transmission_tau20}, together with the result of the non-probabilistic reference solution. We find, that MCRT simulations that use the peel-off method (labeled ``peel-off'') generally lead to better results compared to simulations that use the basic approach (labeled ``original''), which are prone to higher levels of noise and show a stronger underestimation of flux values, typical of simulations in the regime of high optical depths \citep[e.g.,][]{2018ApJ...861...80C,2021A&A...645A.143K}. Furthermore, we find that the SF clearly outperforms the EF for this task, as the estimated transmission is much closer to the reference solution in terms of shape and magnitude. It is important to note that these differences occur despite the fact that both simulations were performed for a fixed number of transmitted PPs. In fact, the EF required $N=157,\!738$ simulated PPs of which only $N_{\rm trans}=10^4$ were transmitted, which is ${\sim}83\,\%$ higher compared to the SF. However, simulation times scale with the number of simulated interactions, $N_{\rm int}$, rather than the number of simulated PPs, which in the case of the EF amount to $N_{\rm int}=6,\!476,\!737$, exceeding the case of the SF by ${\sim}258\,\%$. Consequently, using the SF rather than the EF results in a reduction in the computation time of ${\sim}72\,\%$, while providing better results.

   \begin{figure}
   \centering
   \includegraphics[width=\hsize]{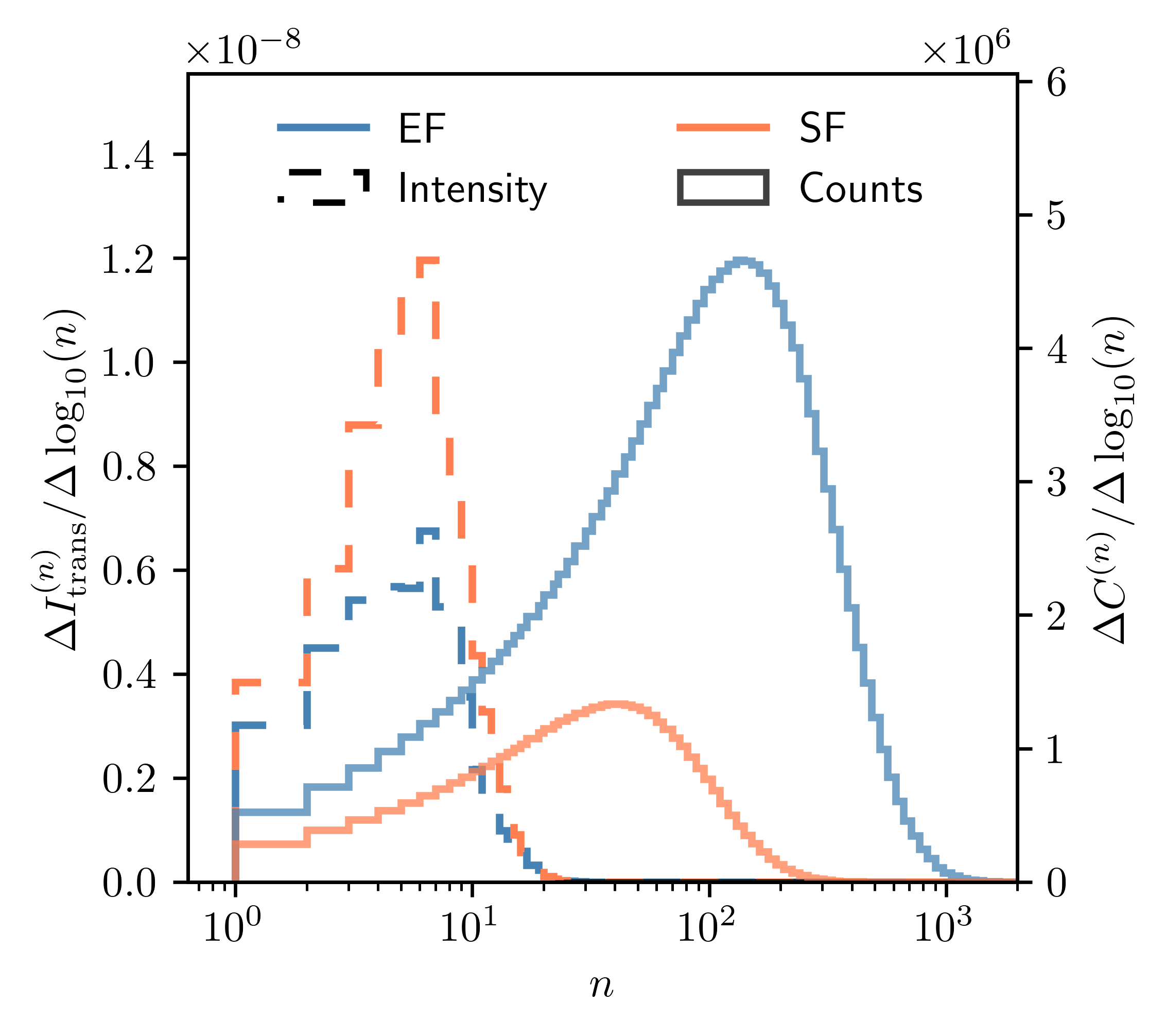}
      \caption{Distributions of counts $C^{(n)}$ of PPs (right axis; solid lines) and corresponding transmitted intensities $I_{\rm trans}^{(n)}$ (left axis; dashed lines) for scattering order, $n,$ measured by the detector bin for direction, $\mu\approx 0$ (perpendicular to the surface of the slab).}
         \label{fig:distributions}
   \end{figure}

    The superiority of the SF over the EF regarding this task can be explained as follows. Since the PP path determination takes place in optical depth coordinates and the albedo satisfies $A<1$, the experienced optical depth of the system is smaller within the SF. In this case, the step length of simulated PP path segments is larger in the SF. As a result, PPs are transported with a faster pace through the model space in terms of traversed units of the extinction optical depth per simulated scattering event. This, in turn, leads to a wider spatial spread of simulated scattering locations, which seems to be crucial for the improvement of flux estimates. Additionally, the complexity of the problem is reduced, indicated by the fact that fewer simulated scattering events lead to better flux estimates. In other words,  switching to the SF has practically lowered the number of PPs that need to be simulated for a sufficiently well-sampled representation of the distribution of scattering orders \citep[SOs;][]{2021A&A...645A.143K}. Here, the SO of a PP describes its total number of  scattering events, $n,$ that have occurred. However, we note in passing that in the limiting case of $A \rightarrow 1$, these differences among both frameworks disappear. 

    \subsection{Impact on scattering order distribution}
    For the detector bin of the direction $\mu\approx 0$, Fig. \ref{fig:distributions} shows the SO dependent number of counts $C^{(n)}$ with the corresponding scattering order distribution (SOD) for detected PPs. The SOD is shown in terms of $\Delta I^{(n)}_{\rm trans} / \Delta \log_{10}(n)$, where $I^{(n)}_{\rm trans}$ is the transmission intensity for SO $n$. These results stem from simulations, in which the peel-off method was applied. As a consequence, the distributions of the number of counts are the same for all detector bins, since during each simulated interaction one peel-off PP is sent to every detector bin. The SODs, however, vary between different detector bins. The plot clearly shows that within the SF less interactions were simulated, which is reflected by the smaller area below the SF-based count distribution. This is especially the case for lower SOs. Despite that, the corresponding SOD exceeds that of the EF, namely, paths that were generated in the SF are overall of greater relevance for the task of determining transmitted intensity values. Moreover, in this framework, PPs often leave the model space with a lower SO, which is beneficial, since the SOs far from the peak of the SOD ($n\gg n_{\rm peak} \approx 6$) barely contribute to the transmission intensity. However, these SOs make up the largest portion of the invested computation time. Comparing the SODs, we find that in the EF the transmission intensity is underestimated across all relevant scattering orders $n>0$, suggesting that the simulation of highly contributing PP paths is less likely in this framework; furthermore, this even fails for the first simulated scattering event. The difference in performance between both frameworks at $\tau_{\rm max}=20$ is striking and can be expected to, firstly, already exist at lower optical depths, and secondly, to increase at higher optical depths. 

    \subsection{Onset of flux underestimation}
    \label{sec:onset_flux_underestimation}
    In the following, we investigate the conditions that lead to the onset of flux underestimation, meaning that we analyze its dependence on the optical depth, albedo, and number of simulated PPs. In contrast to our previous simulations, here we keep the number of PP send out per simulation fixed rather than the number of transmitted PPs. Furthermore, from now on we make use of the peel-off method (unless mentioned otherwise), as it has shown to reliably and significantly boost the quality and performance \citep[e.g., ][]{2021A&A...645A.143K}, see for instance Fig. \ref{fig:transmission_tau20}. 

\begin{figure}
   \centering
   \includegraphics[width=\hsize]{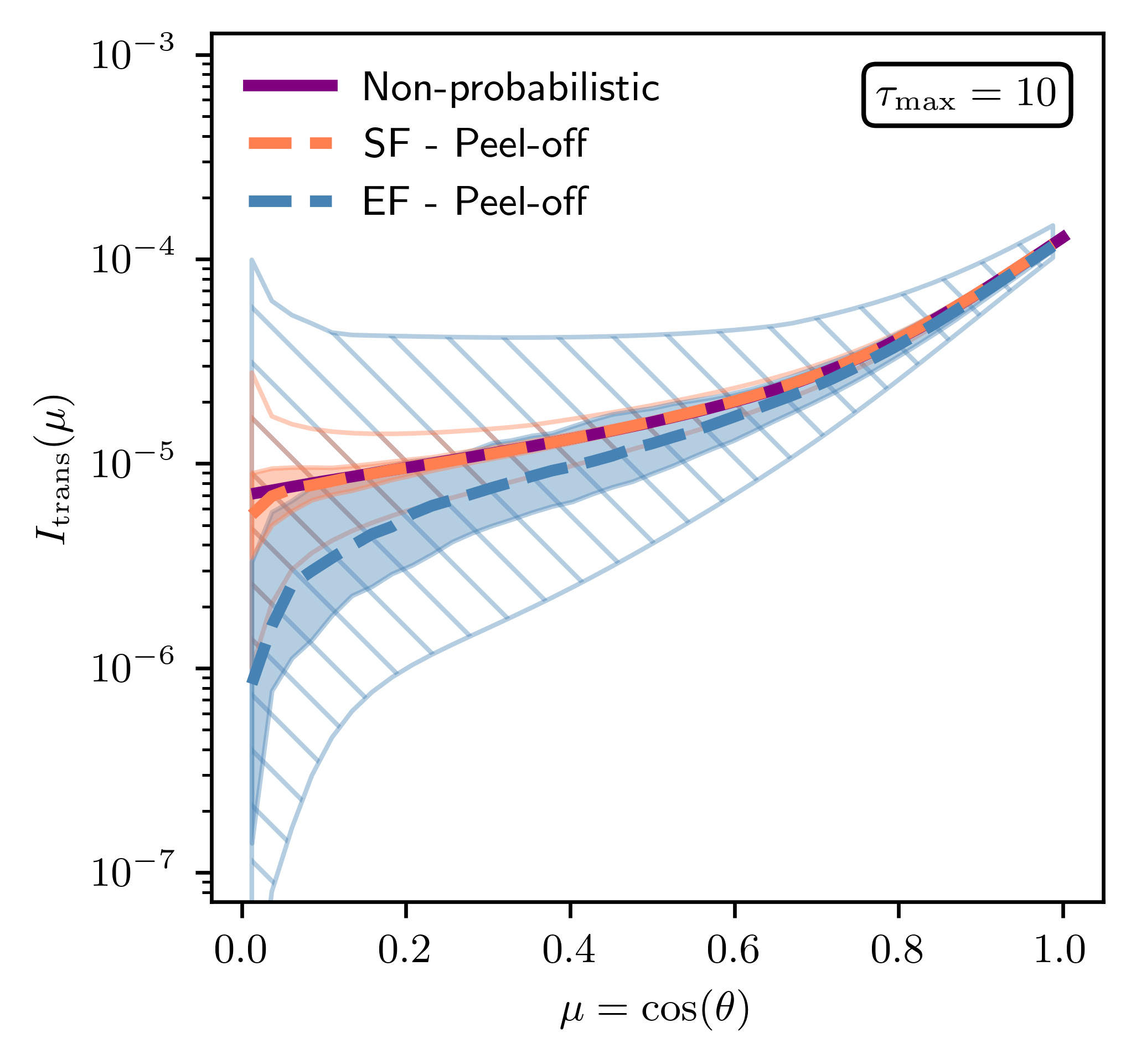}
      \caption{Transmission curves for $\tau_{\rm max}=10$. Results of MCRT simulations comprise median transmission curves, each based on 100 individual simulations using $N=10^4$ PPs. The hatched areas mark regions in the plot within which all 100 simulated transmission curves of the corresponding color lie, while the inner shaded regions display the central 50\% of simulated values.}
         \label{fig:tau10}
   \end{figure}

    For this purpose, we analyze the resulting bias and MC noise of estimates of the transmitted intensity depending on different key parameters: the extinction optical depth $2 \leq \tau_{\rm ext} \leq 20$, the albedo $A \in \left\{0.1,\, 0.5,\, 0.9 \right\}$, and the number of simulated PPs $10 \leq N \leq 10^6$. Additionally, every simulation was repeated 100 times each, allowing for the determination of medians and interquartile ranges (IQRs) to estimate the bias and MC noise of the obtained intensity values, respectively.  
    Figure \ref{fig:tau10} shows the result of the 100 performed simulations assuming $\tau_{\rm max}=10$, $A=0.5$, and $N=10^4$. Dashed lines represent the median transmission intensities for both frameworks, the hatched areas mark the regions in the plot within which all determined transmission curves of the corresponding color lie, and the shaded regions account for the central 50\% of simulated values for each direction, $\mu$. 
    The displayed median curves are suitable for determining the bias of expectable intensity estimates, which are associated with the usage of a limited number of simulated PPs, when performing a MCRT simulation. Similarly, the IQR, which here represents the central 50\,\% of simulated intensity values, can be used as a measure for the expectable spread of obtained intensity estimates caused by the inherent MC noise of the simulations. Comparing the results between the SF and EF, we find that the SF outperforms the EF in the task of determining both with regard to the bias, as its median curve $I_{\rm trans}^{\rm median}$ is a better estimator for the reference solution (purple solid line) and the MC noise (vertical width of shaded area) is significantly smaller for all directions, $\mu$. 
    
    \subsubsection{Dependence on optical depth}
    \label{sec:dependence_optical_depth}
    \begin{figure}
    \centering
    \includegraphics[width=\hsize]{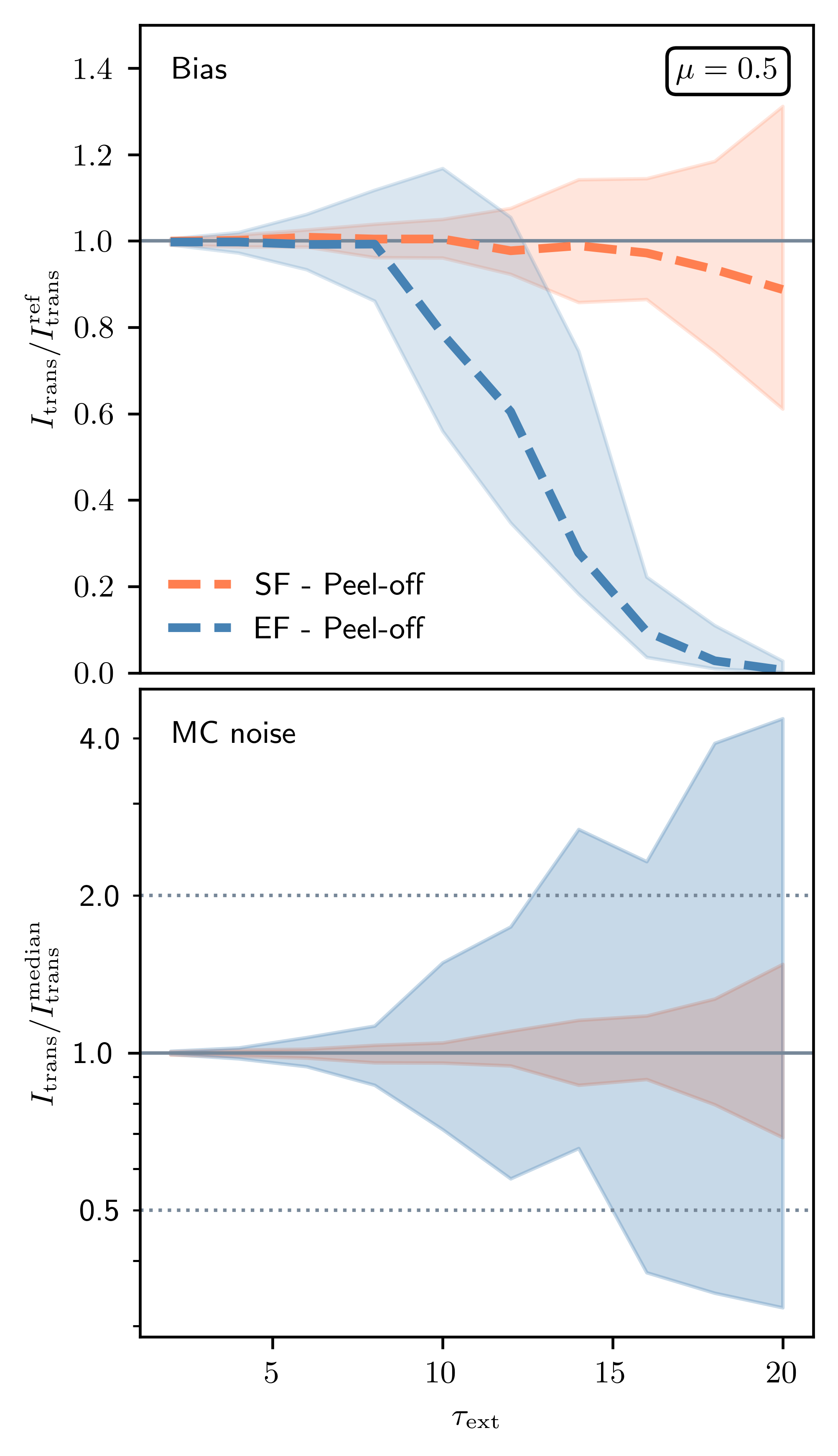}
      \caption{Bias (upper plot) and MC noise (lower plot) as a function of $\tau_{\rm max}$. These results are based on 100 MCRT simulations each, and comprise median transmission curves (dashed lines) using $N=10^4$ PPs and $\mu=0.5$. The shaded regions mark the region within which the central 50\% of simulated values lie. For details, see Sect. \ref{sec:dependence_optical_depth}.}
         \label{fig:tau_dependence}
    \end{figure}
    Furthermore, Fig. \ref{fig:tau_dependence} shows the bias (upper plot) and MC noise (lower plot) as a function of the extinction optical depth of the slab exemplarily for the direction $\mu=0.5$, which corresponds to radiation leaving the slab with a penetration angle of $\theta=60^\circ$. In particular, the bias shows in the plot in the form of underestimated intensity values that lead to $I_{\rm trans}^{\rm median}/I_{\rm trans}^{\rm ref}<1$, where $I_{\rm trans}^{\rm ref}$ represents the (linearly interpolated) reference solution. This plot suggests that for $\tau_{\rm ext} \lesssim 8$ both frameworks lead to a rather small underestimation, however, for higher optical depths, the bias associated with the EF lead to a significant underestimation. On the contrary, we find that the bias of the SF, originating from the limitation of PPs, is much smaller. Hence, it results in significantly better intensity estimates, especially for $\tau_{\rm ext} > 8$, and shifts the onset of the problem toward larger optical depths of $\tau_{\rm ext} \gtrsim 14$. The shape of the displayed shaded areas indicates that the spread of expectable intensity values highly depends on the optical depth. In fact, in the case of the EF, the range of the central 50\,\% of simulated intensity values does not even include the reference values for $\tau_{\rm ext} \gtrsim 12$. To better illustrate the MC noise as a function of the extinction optical depth of the slab, the shaded region in the lower plot in Fig. \ref{fig:tau_dependence} shows the central 50\,\% data points for the ratio $I_{\rm trans}/I_{\rm trans}^{\rm median}$ as a function of $\tau_{\rm ext}$. Here, the vertical width of the shaded region is a measure of the spread of the distribution of intensity estimates. We find, that for increasing optical depth, the noise generally increases in both frameworks. However, this plot clearly suggests that the MC noise associated with the SF is much smaller for all simulated values of $\tau_{\rm ext}$. As a result, it provides a much more reliable estimate for the transmitted intensity than the EF. 

    \subsubsection{Dependence on albedo}
    \label{sec:dependence_albedo}
    We performed the same analysis for a lower ($A=0.1$, Fig. \ref{fig:app:tau_dependence0.1}) and a higher ($A=0.9$, Fig. \ref{fig:app:tau_dependence0.9}) albedo and find that the differences between both frameworks increase for lower values of the albedo. This results in an earlier onset of the problem of intensity underestimation when using the EF. In particular, we find that as low values as $\tau_{\rm ext} > 6$ may lead to errors at the order of $\sim 10\,$\% (see Fig. \ref{fig:app:tau_dependence0.1} in the appendix). Considering the amplification of the bias induced underestimation at lower values of $\mu$ (e.g., see Fig. \ref{fig:tau10}), even higher deviations can be expected. On the contrary, the SF performs reasonably well for all considered optical depth values up to $\tau_{\rm ext}=20$ even with regard to the associated MC noise, which is significantly lower than for the same slab simulations using the EF. Conversely, for $A=0.9$ the performance of both frameworks improves both with regard to the bias and the MC noise (see Fig.  \ref{fig:app:tau_dependence0.1}). For this case, we report only a small benefit from the SF compared to the EF, as can be expected, since both frameworks become more similar as $A\rightarrow 1$.

    \subsubsection{Dependence on number of photon packages}
    \label{sec:dependence_N}
    \begin{figure}
   \centering
   \includegraphics[width=\hsize]{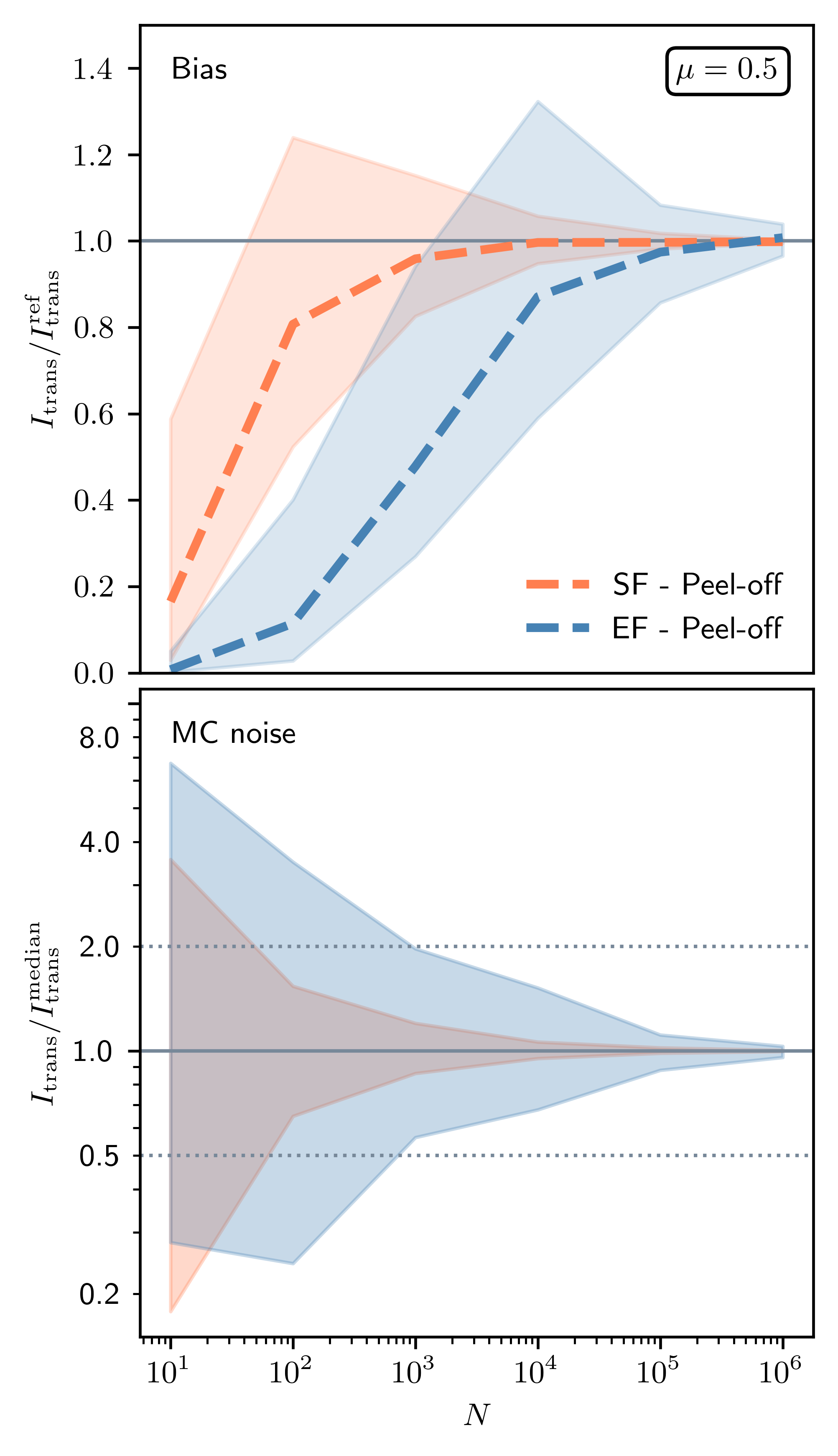}
      \caption{Bias (upper plot) and MC noise (lower plot) as a function of the number of simulated PPs, $N$. These results are based on 100 MCRT simulations each and comprise median transmission curves (dashed lines) using $N=10^4$ PPs and $\mu=0.5$. The shaded regions mark the region within which the central 50\% of simulated values lie. For details, see Sect. \ref{sec:dependence_N}.}
         \label{fig:N_dependence}
   \end{figure}
    Moreover, the number of send out PPs $N$ was varied for a simulated slab of thickness $\tau_{\rm ext}=10$ and its effect on the bias and MC noise of thereby obtained intensity estimates was analyzed. The results of these simulations are displayed in Fig. \ref{fig:N_dependence}, assuming $\mu=0.5$. As expected, the bias in the upper plot decreases for higher values of $N$, which is indicated by the ratio $I_{\rm trans}^{\rm median}/I_{\rm trans}^{\rm ref}$ approaching a value of 1 and, simultaneously, the noise shown in the lower plot decreases. Additionally, the onset of the problem of flux underestimation therefore shifts toward smaller optical depths as $N$ decreases. A comparison of the results of both frameworks clearly suggests that the number of simulated PPs that is required to achieve a low bias is much smaller for the SF than for the EF. In particular, the number of simulated PP that is needed to achieve a median intensity of $\sim 80\,\%$ of the reference solution is about 10 to 100 times higher for the EF than for the SF in this example. Considering the fact that the SF furthermore leads to a smaller number of simulated interactions per PP, this corresponds to a significant potential for saving computation time, already at optical depths as small as $\tau_{\rm ext}=10$, which can be expected to further increase in the regime of higher optical depths.

\section{Discussion and conclusions}
\label{sec:discussion_and_conclusion}

    In the previous section, we compare the results of MCRT simulations that use different frameworks, the EF or the SF, for determining the transmitted intensity through an infinite plane-parallel slab that is illuminated from one side. Our findings clearly suggest that the SF is generally better suited for this task than the EF, as the shape and magnitude of those obtained transmission curves are a better match to the non-probabilistic solution, while at the same time requiring a significantly smaller computation time. The benefits not only entail a smaller bias associated with the usage of the SF, but also greater reliability of transmitted flux estimates due to a smaller MC noise. The observed superiority of the SF may stem from a reduced problem complexity, which seems to be determined in this case by the scattering optical depth rather than the extinction optical depth, with the former being smaller by a factor of $A$.

    Generally, the albedo plays a crucial role in determining the complexity of a MCRT problem. For a source that is deeply embedded inside a region of high optical depth, for instance, the number of interaction events scales approximately quadratically with the optical depth \citep[e.g.][]{2020A&A...635A.148K,2021A&A...645A.143K}. In the SF, the computation time can therefore be expected to also scale with $\sim A^2$ when simulating deeply embedded sources of radiation (e.g., protostars, protoplanets, visous heating in a protoplanetary disk, etc.). 
    Considering the results for the SODs shown in Fig. \ref{fig:distributions}, the improved quality manifests itself for all $n>0$, suggesting, that the SF is promoting highly contributing paths at a much higher rate than the EF. Moreover, it is known that for higher values of $A$; firstly, the SOD shifts toward larger scattering orders and, secondly, broader SO intervals have to be simulated to ensure a reliable transmitted intensity estimation \cite{2021A&A...645A.143K}. The path determination in the SF, interestingly, follows exactly these trends, while the EF path determination utilizes a procedure that is completely independent of the albedo. Therefore, it can be expected that the SF outperforms the EF for all values of $A$, which is supported by our results (presented in Sect. \ref{sec:results}).

    Our analysis of the onset of flux underestimation (Sect. \ref{sec:onset_flux_underestimation}) has shown that this issue can manifest itself already at optical depths as small as $\tau_{\rm ext} \approx 6$ or lower, depending on the number of feasibly simulated PPs per source. Such an early onset of this problem may very well explain the high level of MC noise present in MCRT simulations of embedded planets \citep[][]{2022A&A...662A..99K} or of the process of self-scattering \citep[e.g.,][]{2020A&A...640A.122B}, which were both performed within the EF. Meaning, a shift to the SF may significantly benefit such simulations in the form of a reduced simulation times and less MC noise. 

    To evaluate the performance of MCRT simulations in the regime of high optical depths, we additionally performed simulations for a slab of transverse optical depth $\tau_{\rm max}=75$ for $A \in \left\{0.1,\,0.5,\,0.9 \right\}$ and determined the transmitted intensity (see Fig. \ref{fig:transmission_tau75}). Here, we analyzed the interplay between the framework, the albedo, and an additional usage of composite-biasing \citep{2016A&A...590A..55B}, a prominent method that affects the path determination. These simulations provided further support for the thus far presented conclusions (for details, see Sect. \ref{sec:appendix:high_optical_depths}). Overall, using the SF resulted in benefits for all tested albedos and optical depths, both independent of the additional usage of the composite-biasing method.

    Despite the benefits the SF clearly offers, there are cases, in which the EF may outperform the SF. For instance, the back-scattered part of radiation may be misrepresented in the optically thick case when $A\ll 1$. In this case, the corresponding flux values may be underestimated as a consequence of the less densely sampled distribution of scattering locations close to the illuminated side of the slab. Whether this is of concern, however, depends on the particular type of simulation and the specifics of the simulated system. In general, when deciding on the framework, it seems crucial to consider the distribution of simulated interaction points. If, for instance, a dense distribution close to the origin of radiation (in units of optical depth) is desired, the EF may be beneficial. Similarly, if the optical depth is very small, the EF may have benefits due to a denser sampling; however, this could be compensated within the SF by the usage of a method that enforces a minimum number of simulated scattering events per PP. A dedicated study that aims at addressing these topics would certainly be of great interest. 

    However, apart from the path determination procedure, both frameworks also exhibit striking differences regarding their weight determination schemes, which has implications with respect to their benefits and shortcomings. In general, the contribution of a simulated PP path to a flux estimate equals the product of its probability (density) and weight. Considering that the means by which a deviation from some most contributing path is penalized by these frameworks differs. In particular, the SF penalizes a deviation by a weight associated with the spatial deviation from the most contributing path. The EF, on the other hand, penalizes by attributing a lower probability for traversing large extinction optical depths. In the EF, there is always a residual possibility for a simulated PP to traverse any optical depth without losing weight, while in the SF the weight always decreases as the PP traverses an absorbing medium. Therefore, a PP transmitted within the SF always carries information about its complete previous path, as its total length passively affected its (energy) state due to the application of a corresponding weighting factor. As a result, the final weight may vary significantly stronger within the EF, resulting in an overall higher MC noise of the flux estimates (see Sect. \ref{sec:results}) even at low optical depths, and notably in the regime of high optical depths. As a result, choosing the SF over the EF likely results in a variance reduction, which has been supported by our results. In other words, "passifying" the absorption process allows us to reduce the number of simulated scattering events while simultaneously decreasing the bias and inherent MC noise. This raises the question of whether it is possible to passify other probabilistic procedures in MCRT simulations or improve them by utilizing machine learning techniques such as reinforcement learning to generate probability density functions for random variables that depend on the simulated environment. 
        
    Based on the results of this study, we can conclude that the SF clearly outperforms the EF in the task of determining transmission intensities using MCRT simulations. We showed that this affects the quality of the results significantly in the form of a reduced bias, smaller MC noise, and an alleviation of the problem of underestimated flux values. Additionally, the computational demand to achieve a low bias within the SF can be multiple orders of magnitude lower, leading to an overall considerably faster convergence speed. Its implementation can therefore greatly benefit all modern MCRT codes that are aimed at determining transmitted flux values.

\begin{acknowledgements}
    We thank all the members of the Astrophysics Department Kiel for helpful discussions and remarks. 
    We acknowledge the support of the DFG priority program SPP 1992 "Exploring the Diversity of Extrasolar Planets (WO 857/17-2)".
\end{acknowledgements}

\bibliography{literature}
\bibliographystyle{aa}

\begin{appendix}
\section{Framework transformation}
\label{sec:appendix:EF2SF}
    In the following, we demonstrate how the EF and the SF are linked via the application of a biasing technique. By doing so, we furthermore show, that the discrete nature of removing PPs at the location of scattering in the EF thereby transforms into the continuous absorption scheme that is applied in the SF. In the EF, the weighting factor is $w=A$ and the traversed extinction optical depth is selected by the determination of a pseudo random number $r\in \left[0,\,1\right)$ according to a uniform distribution, followed by its calculation: 
    \begin{equation}
        \tau_{\rm ext}=-\ln{\left(1-r\right)}. 
        \label{eq:app:standard_tauext}
    \end{equation}
    Therefore, the probability density function for the random variable, $\tau_{\rm ext}$, is given by ${\rm d}r/{\rm d}\tau_{\rm ext} = \exp\left(-\tau_{\rm ext} \right)$. Next, we assume that a PP is emitted and follows a straight path until it interacts at a distance $l\in \left[l^*, l^*+ \delta l \right)$ with the medium, where $\delta l \ll l^*$ is small. In the EF, this interval corresponds to an interval of random numbers $r \in \left[r_i, r_i+ \delta r \right)$ and to an optical depth interval $\tau_{\rm ext}\in \left[\tau_e, \tau_e+ \delta \tau_e \right)$, where $\tau_e = l^* \rho\, C_{\rm ext}$ and $\delta \tau_e = \delta l\, \rho\, C_{\rm ext}$. The expected portion of its initial energy $f_{\rm sca}$, that is scattered within this region, is then given by
        \begin{eqnarray}
            f_{\rm sca} &=& \int_{r_i}^{r_i+ \delta r} {\rm d}r\, A, \\
            &=& \int_{\tau_e}^{\tau_e + \delta \tau_e} {\rm d}\tau_{\rm ext}\, \left(\frac{{\rm d}r}{{\rm d}\tau_{\rm ext}}\right) A, \\
            &=& \int_{\tau_e}^{\tau_e + \delta \tau_e} {\rm d}\tau_{\rm ext}\, e^{-\tau_{\rm ext}} A, 
            \label{eq:app:ef_calc} \\
            &=& A \delta \tau_e e^{-\tau_e}.
        \end{eqnarray}
    Similarly, in the SF, the interval corresponds to an optical depth interval $\tau_{\rm sca}\in \left[\tau_s, \tau_s+ \delta \tau_s \right)$, where $\tau_s = l^* \rho\, C_{\rm sca} = A\tau_e$ and $\delta \tau_s = \delta l\, \rho\, C_{\rm sca} = A \delta \tau_e$. In this framework, the scattered portion of energy thus equals:
    \begin{equation}
        f_{\rm sca} = \int_{\tau_s}^{\tau_s + \delta \tau_s} {\rm d}\tau_{\rm sca}\, e^{-\tau_{\rm sca}} e^{-\tau_{\rm abs}},
        \label{eq:app:sf_term}
    \end{equation}
    where $e^{-\tau_{\rm abs}}$ is the weighting factor accounting for the continuous absorption of energy, which eventually yields the same result as the calculation in the EF. 

    Next, we assume that the determined PP path is stretched in Eq. \eqref{eq:app:standard_tauext} by a constant factor of $\alpha$, namely, $\tau_{\rm ext}=-\alpha \ln{\left(1-r\right)}$, such that:
    \begin{equation}
        \left(\frac{{\rm d}r}{{\rm d}\tau_{\rm ext}}\right) = \frac{1}{\alpha} e^{-\tau_{\rm ext}/\alpha}.
        \label{eq:app:q19-diff}
    \end{equation}
    We can then use this result and re-write Eq. \eqref{eq:app:ef_calc} as follows:
    \begin{eqnarray}
        &&\int_{\tau_e}^{\tau_e + \delta \tau_e} {\rm d}\tau_{\rm ext}\, e^{-\tau_{\rm ext}} A, \\
        &=& \int_{\tau_e}^{\tau_e + \delta \tau_e} {\rm d}\tau_{\rm ext}\, 
        \frac{1}{\alpha} e^{-\tau_{\rm ext}/\alpha}
        \left( \alpha e^{-\tau_{\rm ext}+\tau_{\rm ext}/\alpha} A \right),
    \end{eqnarray}
    where the term in front of the bracket corresponds to the probability density function for stretched PP paths and the term inside the bracket is the resulting corresponding weighting factor. 
    When switching from the EF to the SF, the stretching factor is $\alpha = 1/A$, which, when plugging it into the previous equation, yields
    \begin{eqnarray}
        && \int_{\tau_e}^{\tau_e + \delta \tau_e} {\rm d}\tau_{\rm ext}\, e^{-\tau_{\rm ext}}  \\
        &=& \int_{\tau_e}^{\tau_e + \delta \tau_e} {\rm d}\tau_{\rm ext}\, 
        A e^{-\tau_{\rm ext} A}
        \left( e^{-\tau_{\rm ext}+\tau_{\rm ext} A} \right)  \\
        &=& \int_{\tau_s}^{\tau_s + \delta \tau_s} {\rm d}\tau_{\rm sca}\, e^{-\tau_{\rm sca}} e^{-\tau_{\rm abs}}.
        \label{eq:app:q19-sf}
    \end{eqnarray}
    Here, we used a coordinate transformation and substituted $\tau_{\rm ext}$ with $\tau_{\rm sca}/A$ to arrive at Eq. \ref{eq:app:q19-sf}. As can be seen, this is the same expression as used in the SF (see Eq. \eqref{eq:app:sf_term}). Meaning, by selecting $\tau_{\rm ext}$ values and stretching them to match $\tau_{\rm sca}$, it is necessary to apply the weighting factor $w=e^{-\tau_{\rm abs}}$, which can be interpreted as the continuous absorption of photons along the path. As a result, the EF, in which no stretching is applied (i.e., $\alpha=1$), is the only case in which the weighting factor is independent of the selected path length and there is no continuous change of energy along the path.

\section{Regime of high optical depths}
\label{sec:appendix:high_optical_depths}
    In order to evaluate the performance of MCRT simulations in the regime of high optical depths, we additionally performed simulations for a slab of transverse optical depth of $\tau_{\rm max}=75$ and albedo $A=0.5$, and determined the transmitted intensity, using $N=10^6$ send out PPs. However, since the obtained transmission curves exhibit a high level of noise, which is a consequence of the high optical depth, each MC-based simulation was repeated 100 times. Apart from the calculation of a non-probabilistic solution for the problem, we performed three types of MC-based simulations, which all included the peel-off method. One simulation type is based on the SF and does not apply any further MCRT methods (besides the peel-off method) to boost its performance. The other two types additionally use composite-biasing \citep{2016A&A...590A..55B}, which is a MCRT biasing technique used for stretching PP paths and both types differ regarding their utilized frameworks. Stretching paths is particularly interesting, as it has the potential to greatly improve transmitted flux estimates in the regime of high optical depths, \citep{2018ApJ...861...80C,2021A&A...645A.143K} and its performance has not previously  been tested in the SF. 
    
\begin{figure}
   \centering
   \includegraphics[width=\hsize]{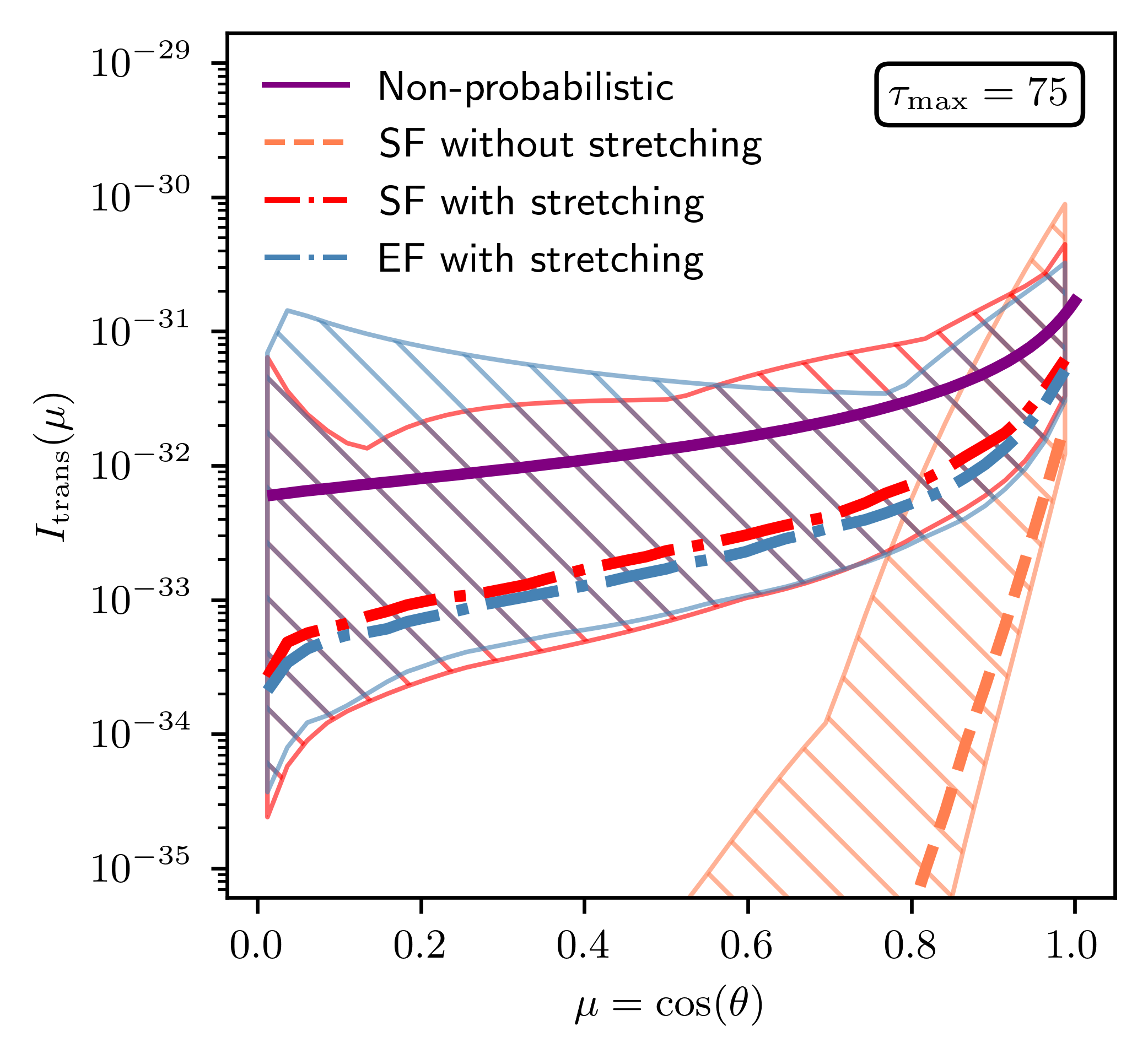}
      \caption{Transmission for $\tau_{\rm max}=75$. MC-based results comprise median transmission curves, each based on 100 peel-off-based individual simulations with $N=10^6$ send out PPs per simulation. The hatched areas mark regions in the plot within which all 100 simulated transmission curves of the corresponding color lie. For details, see Sect. \ref{sec:appendix:high_optical_depths}.}
         \label{fig:transmission_tau75}
   \end{figure}

    The results of these in total 300 MC-based simulations and the non-probabilistic solution are shown in Fig. \ref{fig:transmission_tau75}. We find that simulations that do not use the stretching method fail to properly estimate the transmission curve for all penetration directions, $\mu$. Therefore, using the SF does not by itself suffice to solve the problem of high optical depths, and further techniques are needed to efficiently sample highly contributing paths. When combining the stretching method with either of the two frameworks, the results significantly improve and both types of simulations seem to perform similarly well. This suggests that the usage of the composite-biasing method has practically alleviated the downsides of the EF enough, such that the performance of both frameworks appears to be equivalent. However, it must be stressed, that this is found to be the case for an albedo of $A=0.5$. 
    
    In further tests, we found that the performance of these three types of simulations strongly depends on the albedo. In the case of a small albedo ($A=0.1$), it seems that switching to the SF suffices to properly estimate the transmission, such that no stretching method was needed. Using the EF without stretching, however, resulted in significantly underestimated intensities. This may be explained by the fact that the scattering optical depth of the system equals only $7.5$, which is not particularly high. Therefore, the complexity of the MCRT simulations within the SF is much lower compared to the case of EF-based simulations. For an albedo of $A=0.5$, the SF without stretching has shown to improve the quality beyond similar simulations performed in the EF, which also further increases when additionally using the composite-biasing technique. Finally, if the albedo is relatively large ($A=0.9$), the SF still performs better than the EF when stretching is not used. However, the reduction of computation time was not particularly high in this case. As expected, though, we find that using the composite-biasing method degrades the quality of intensity estimates, which is likely caused by an insufficient coverage of the underlying SOD, in accordance with the findings of \citet{2021A&A...645A.143K}.

    In summary, we conclude that using the SF rather than the EF resulted in overall lower computation times and consistently better transmitted intensity estimates. This is independent of the fact of whether the stretching method was applied or not.

\section{Supplementary material}
\label{sec:appendix:supplementary_material}
    \begin{figure}
    \centering
    \includegraphics[width=\hsize]{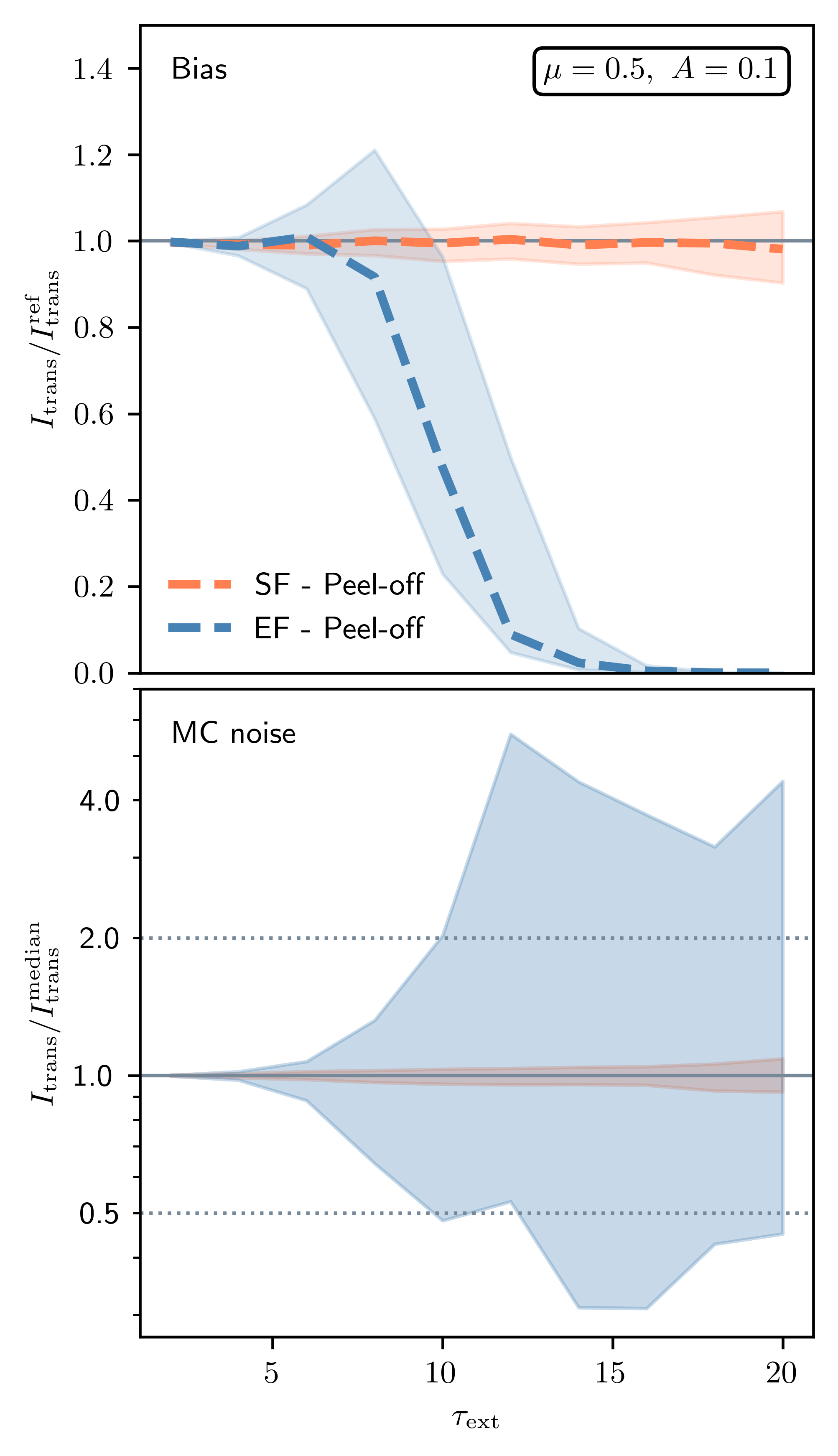}
      \caption{Similar to Fig. \ref{fig:tau_dependence}, but assuming $A=0.1$. For details, see Sect. \ref{sec:dependence_albedo}.}
         \label{fig:app:tau_dependence0.1}
    \end{figure}

    \begin{figure}
    \centering
    \includegraphics[width=\hsize]{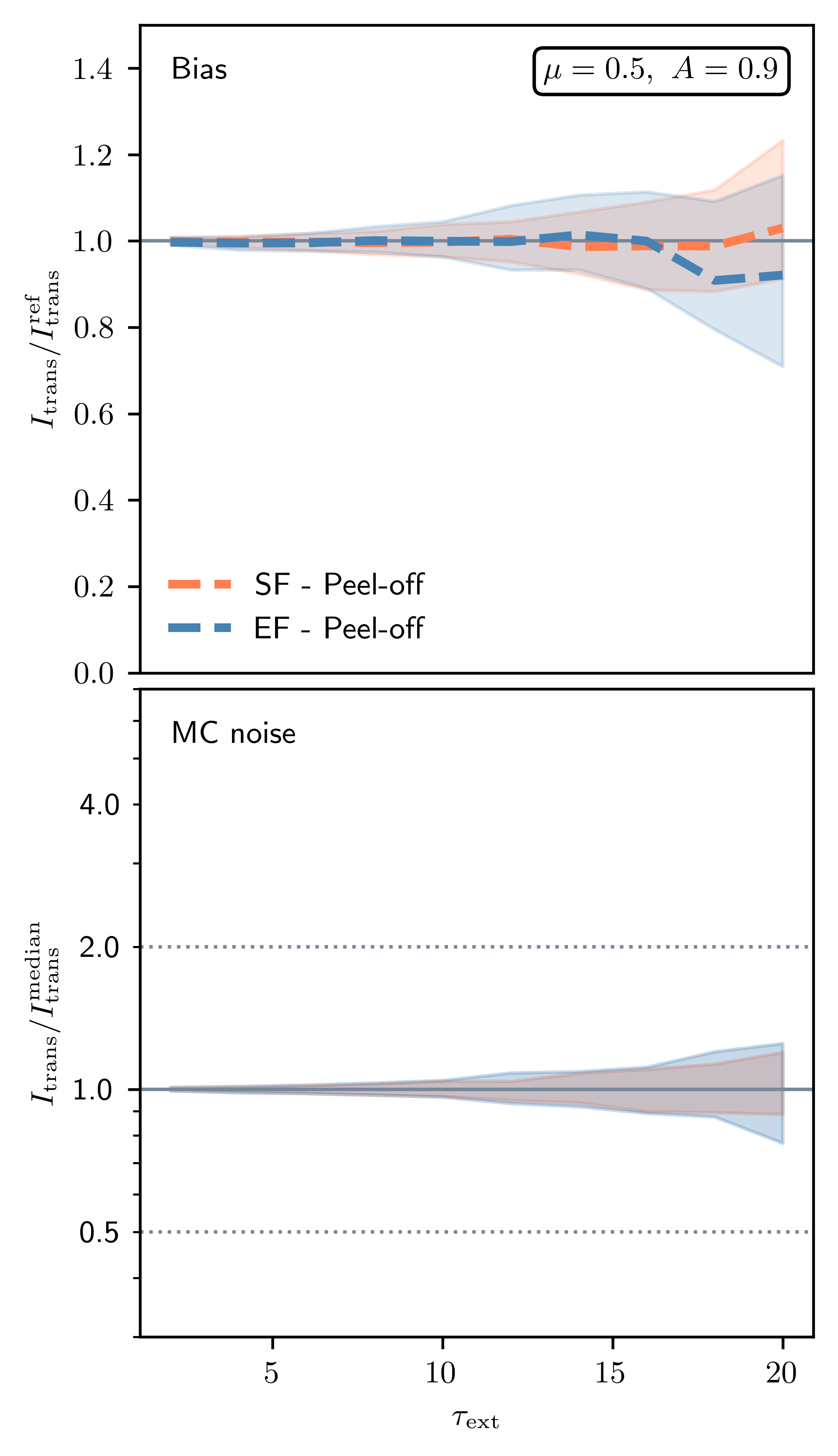}
      \caption{Similar to Fig. \ref{fig:tau_dependence}, but assuming $A=0.9$. For details, see Sect. \ref{sec:dependence_albedo}.}
         \label{fig:app:tau_dependence0.9}
    \end{figure}

\end{appendix}

\end{document}